\documentclass[%
reprint,
superscriptaddress,
twocolumn,
amsmath,amssymb,
aps,
pre,
]{revtex4-2}

\usepackage{amssymb,amsmath,bm}
\usepackage{enumerate}
\usepackage[utf8]{inputenc}
\usepackage{cancel}
\usepackage{ulem}
\usepackage{color}
\usepackage[colorlinks,linkcolor=blue,urlcolor=blue,citecolor=blue]{hyperref}
\usepackage{float}
\usepackage{xr}
\usepackage[toc]{appendix}
\usepackage{graphicx,epstopdf}
\usepackage{tikz}

\usepackage{mathabx}

\def\be{\begin{equation}}
	\def\ee{\end{equation}}
\def\bea{\begin{eqnarray}}
	\def\eea{\end{eqnarray}}

\begin{document}
\title{Thermodynamic criticality of coupled oscillators}
\author{Suvam Pal}
\email{suvamjoy256@gmail.com}
\affiliation{Physics and Applied Mathematics Unit, Indian Statistical Institute, 203 B.T. Road, Kolkata, India}

\author{Sudipta Mukherji}
\email{mukherji@iopb.res.in}
\affiliation{Physics and Applied Mathematics Unit, Indian Statistical Institute, 203 B.T. Road, Kolkata, India}

\author{J\"urgen Kurths}\email{kurths@pik-potsdam.de}
\affiliation{Potsdam Institute for Climate Impact Research, Potsdam, Germany}
\affiliation{Department of Physics, Humboldt-Universit\"at zu Berlin, Berlin, Germany}
\affiliation{Research Institute of Intelligent Complex Systems, Fudan University,
Shanghai, China}

\author{Dibakar Ghosh}
\email{Corresponding author: dibakar@isical.ac.in}
\affiliation{Physics and Applied Mathematics Unit, Indian Statistical Institute, 203 B.T. Road, Kolkata, India}

\begin{abstract}
      Strict thermodynamic scaling relations, such as the Rushbrooke
    inequality, are fundamentally established for equilibrium critical
    phenomena in the thermodynamic limit. In finite-size dynamical systems exhibiting synchronization, the direct application of such identities is hindered both by the finiteness of the network and the nonequilibrium nature of the spontaneous synchronization transition. To bypass this difficulty, we rigorously study the dynamical counterparts of the order parameter, susceptibility, and specific heat in finite systems of dynamical oscillators with nonlinear coupling. By measuring these quantities as a function of system size, we extract the associated critical exponents governing the transition. The validity of our thermodynamic mapping is tested by directly confirming the Rushbrooke inequality. Our results establish that standard equilibrium thermodynamic scaling architectures can be systematically applied to the finite-size scaling of nonequilibrium synchronization dynamics.
\end{abstract}

\maketitle

\textbf{\textit{Introduction}}--
The emergence of collective behavior in interacting dynamical systems has remained in the limelight over the past few decades. Ranging from Japanese tree frogs to neuronal networks, almost every field of science exhibits such behavior. Depending on the topological variance \cite{kuramoto1975international,kuramoto1984chemical,watts1998collective,yeung1999time}, interaction strength \cite{strogatz2000kuramoto}, or the dimensionality of the systems, a sharp transition from coherent to incoherent state is encoded. Wherein, the thermodynamic limit--where the number of degrees of freedom $N$ approaches infinity--provides the mathematical tractability necessary to rigorously define these sharp transitions~\cite{yeung1999time,strogatz2000kuramoto}. Yet, physical reality dictates that natural and engineered networks~\cite{strogatz2005crowd,antzoulatos2014increases,nixon2013observing,motter2013spontaneous,pikovsky1985universal}, from biological swarms~\cite{winfree1967biological,o2017oscillators} to coupled physical arrays~\cite{zhang2015explosive,hong2015finite}, possess inherently finite populations.

The infinite-size idealization is a powerful analytical tool. However, the finite nature of real systems is not merely a source of secondary quantitative error, but rather a primary qualitative driver of the underlying dynamics. In finite oscillator networks~\cite{lee2014finite,PhysRevE.110.034216}, dimensional constraints actively amplify intrinsic fluctuations. Rather than simply shifting the critical point~\cite{kuehn2021universal}, these fluctuations profoundly destabilize macroscopic coherence, effectively smearing the sharp phase transitions canonically predicted by thermodynamic approximations.

In this work, we investigate a precise crossover regime, where these finite-size fluctuations dominate the collective behavior of the interacting dynamical degrees of freedom. We consider a globally coupled system of oscillators with finite ensemble size, driven by noise and a symmetry-breaking external field. Notably, in the dynamical perspective, this field is known as pinning. This drive acts analogously to an aligning magnetic field in the classical Ising~\cite{bruinsma1984interface, binder2002monte, tissier2011supersymmetry} or XY~\cite{casetti1999topological,hu2021extraordinary} models, allowing us to capture the exact mechanism by which finite populations degrade macroscopic order.

Considering microscopic coupled Langevin equations, we derive the exact stationary state probability density function via the corresponding Fokker-Planck formalism. By analyzing the system’s response to the external drive, we extract the associated critical exponents $(\alpha, \beta, \gamma)$ that govern various scaling relations. Crucially, our analytical derivations and rigorous numerical simulations validate the Rushbrooke inequality $(\alpha + 2\beta + \gamma \ge 2)$. In critical phenomenon, this inequality arises from the scaling hypothesis. This hypothesis states that, in the vicinity of a critical point, the singular part of the thermodynamic free
energy is a generalized homogeneous function of its controlling
parameters. Our analysis, therefore, provides a rigorous mathematical
bridge structurally linking the non-equilibrium transition of finite
oscillator networks with the established framework of equilibrium
statistical phenomena.


\begin{figure}
    \centering
    \includegraphics[width=\linewidth]{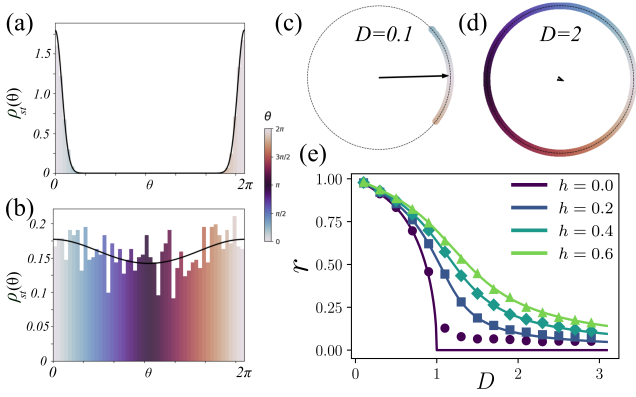}
    \caption{Schematic representation of microscopic behavior of the phase and variation of order-parameter $r$ with respect to noise-strength $D$. Panels (a) and (b) illustrate the probability density profile of phases ($\{\theta\}$) coherent and incoherent states at $t\to \infty$, respectively, in the absence of external field $h$. The corresponding configurations of the phases are represented in panels (c) and (d). Panel (e) describes the variation of $r$ as a function of $D$ for different external field strengths $h$. Solid lines indicate the theoretical estimate (see Eq.~\eqref{r}), and markers indicate the numerical estimate.}
    \label{fig1}
\end{figure}

\textbf{\textit{Model setup}}--
Synchronization is one of the widely observed phenomena in every scale of physical and engineered systems, such as neuronal networks and epilepsy in the brain, circadian rhythm, the collapse of the Millennium Bridge, and the collective behavior of swarming robots, power grids, and social behavior of humans~\cite{pikovsky1985universal}. To investigate the critical phenomena underlying these collective behaviors, we begin by considering an ensemble of globally coupled oscillators, induced with an external field of strength $h$.

The dynamical evolution of the ensemble of oscillator phases, denoted with $\theta_i$ for $i=1,2,\cdots,N$, is governed by the $N$ coupled Langevin equations
\begin{equation}\label{dynamical-eq}
    \dot{\theta}_i=-h\sin\theta_i+K\mathcal{F}(\Phi(\{\theta\})-\theta_i)+\xi_i(t),
\end{equation}
where $K$ represents coupling strength and $\xi_i(t)$ encodes the dynamical heterogeneity in terms of thermal noise with strength $D$, satisfying $\langle\xi_i(t)\rangle=0$ and $\langle \xi_i(t)\xi_j(s)\rangle=2D\delta_{ij}\delta(t-s)$. The external drive, $-h\sin\theta_i$, acts analogously to an aligning field in the Ising or XY model, energetically favoring the pinning of individual phases near $\theta=0$, as illustrated their density in Figs.~\ref{fig1}(a,b) and corresponding state configurations in Figs.~\ref{fig1}(c,d) respectively. In the second term of the right-hand side in Eq.~\ref{dynamical-eq} $\Phi\{\theta\}$ indicates a collective effect of all phases on a particular phase $\theta_i$. Assuming standard mean-field coupling, the interaction simplifies to $\mathcal{F}(\Phi(\{\theta\})-\theta)\sim r\sin(\psi-\theta)$, where the macroscopic complex order parameter is defined as $z=re^{\iota \psi}=\sum_{j=1}^Ne^{\iota \theta_j}/N$ with the amplitude $r\in [0,1]$. Note that $r\to 0$ and $r\to 1$ measure the correlation in incoherent and coherent states, respectively. Under this formulation, the microscopic dynamics (Eq.~\eqref{dynamical-eq}) is reduced to
\begin{equation}
    \dot{\theta}_i=-h\sin\theta_i+Kr\sin(\psi-\theta_i)+\xi_i(t).
\end{equation}
In the presence of an external field, the mean phase $\psi$ naturally aligns with $0$. In the zero-field limit ($h\to 0$), the rotational symmetry allows us to use $\psi=0$ without loss of generality. Taking the thermodynamic limit ($N\to \infty$), the single-oscillator dynamics is accurately described by
\begin{equation}\label{thermo-dyn}
    \dot{\theta}_i=-h\sin\theta_i-Kr\sin\theta_i+\xi_i.
\end{equation}
Defining $\rho(\theta,t)$ as the probability density function (PDF) to characterize the configuration at any time $t$, the temporal evolution is governed by the corresponding  Fokker-Planck equation
\begin{equation}\label{fp}
    \dfrac{\partial }{\partial t}\rho(\theta,t) = \dfrac{\partial}{\partial \theta}\bigg[(h+Kr)\sin\theta~\rho(\theta,t)\bigg]+D\dfrac{\partial^2 }{\partial \theta^2}\rho(\theta,t).
\end{equation}
At $t\to \infty$, the system relaxes to a stationary state, which yields
\begin{equation}\label{stationary}
    \rho_{st}(\theta)=\dfrac{1}{\mathcal{Z}}\exp\left[\dfrac{h+Kr}{D}\cos\theta\right],
\end{equation}
with the normalizing factor $\mathcal{Z}=2\pi I_0((h+Kr)/D)$, where $I_0(x)$ represents the modified Bessel function of the first kind of order $0$. In the absence of the external field $(h=0)$, the stationary states (see Eq.~\eqref{stationary}) are illustrated in Figs.~\ref{fig1} (a) and (b) of the coherent and incoherent states, respectively. Notably, in the zero-noise limit ($D\to 0$), the $\rho_{st}(\theta)$  naturally collapses to a Dirac delta function $\delta(\theta)$ (see \textbf{Eq.~(S10) in SM}). 

The stationary order parameter $r$ is evaluated via $r=\mathbb{R}\left[\int_0^{2\pi} \rho_{st}(\theta) e^{\iota \theta} d\theta\right]$, yielding the exact self-consistent equation
\begin{equation}\label{r}
    r=\left.I_1\left(\dfrac{h+Kr}{D}\right)\right/I_0\left(\dfrac{h+Kr}{D}\right),
\end{equation}
where $I_1(x)$ represents the modified Bessel function of the first kind of order $1$.
\begin{figure}
    \centering
    \includegraphics[width=\linewidth]{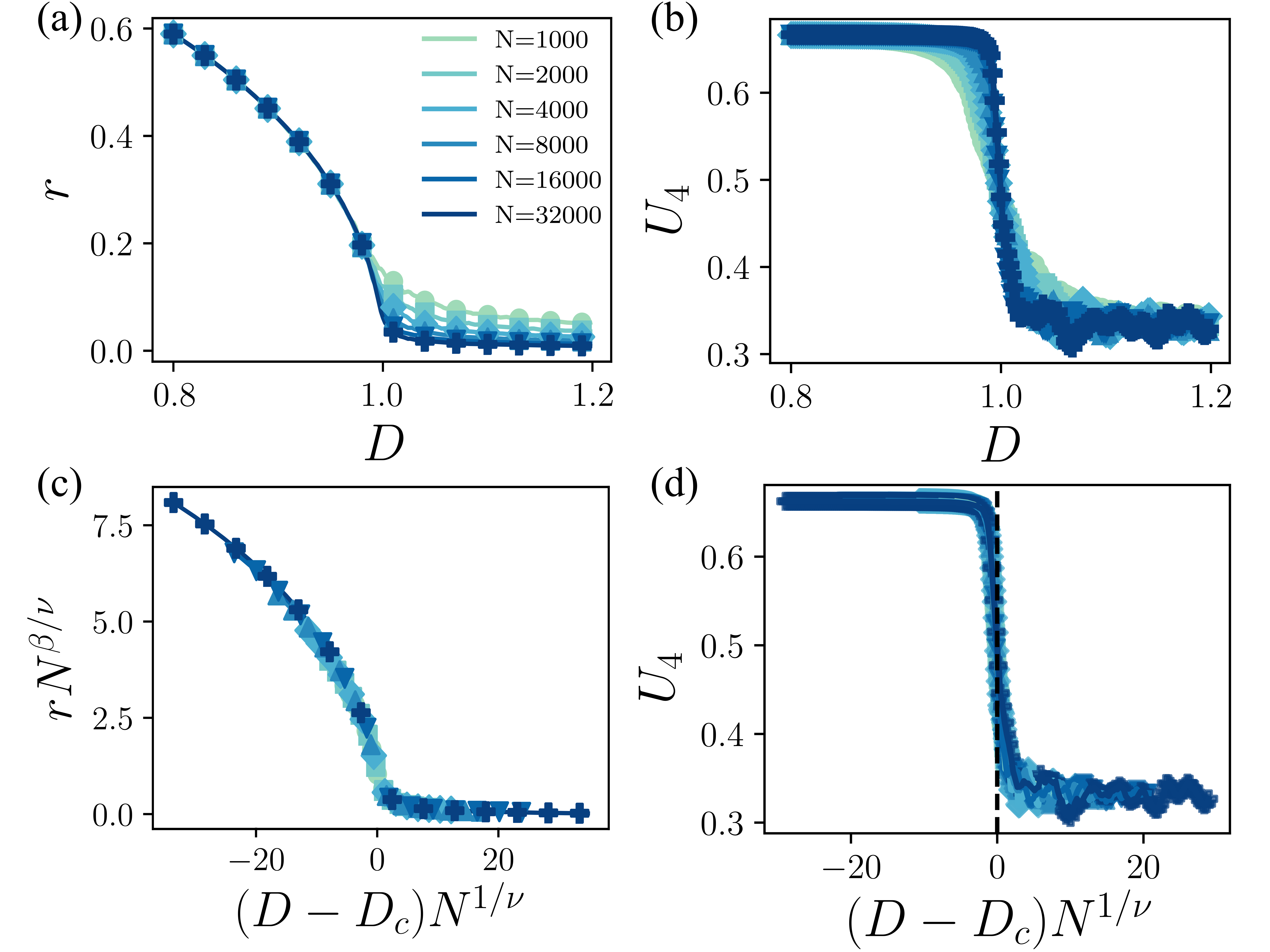}
    \caption{Finite-size scaling of the synchronization transition in the absence of external field ($h=0$). Panel (a) illustrates the variation of the $r$ by calibrating $D$ with different sets of ensemble sizes ranging from $N=10^3$ to $32\times 10^3$. Finite-size effects severely smear the transition near the critical point. The variation of Binder cumulant $U_4$ with $D$ is represented in panel (c). The invariant crossing points accurately pinpoint the critical noise strength $D_c\simeq 0.999$ for a fixed coupling strength $K=2$. In panel (b) and (d), we showcase the data collapse for $r$ and $U_4$, respectively, achieved by anchoring the correlation length exponent at $\nu=2$. This collapse yields the order parameter critical exponent to be $\beta=0.496\pm 0.011$.}
    \label{fig2}
\end{figure}
In the thermodynamic limit with zero external drive ($h=0$), $r$ undergoes a continuous phase transition from the coherent to incoherent state at a critical noise strength $D_c$. A non-zero external field critically alters this behavior, breaking rotational symmetry and inducing residual coherence beyond $D_c$ as indicated in Fig.~\ref{fig1}(e). The primary focus of this article is to characterize the finite-size scaling behavior in the critical region near $D_c$.

\textbf{\textit{Scaling near criticality}}--To explore this, we numerically evaluate $r$ as a function of $D$ for various size of ensembles $N$ at $h=0$. As described in Fig.~\ref{fig2}(a), the finite-size effect severely smears the transition. To precisely locate the critical point independent of this finite-size smearing, we analyze the Binder cumulant $U_4$ of the order parameter $r$; mathematically, this is defined as $1-\frac{\langle r^4\rangle}{3\langle r^2\rangle^2}$, where $\langle \cdots\rangle$ indicates the ensemble average of a quantity. The invariance crossing points of $U_4$ accurately pinpoint the critical noise strength $D_c\simeq 0.999$ (see Fig.~\ref{fig2}(c)), while the coupling strength $K$ is fixed at $2$. Notably, this indicates the consistency with the theoretical prediction $D_c=K/2$. Generally, the order-parameter $r$ follows a power-law behavior near the criticality with an exponent $\beta$, that reads
\begin{equation}
    r\approx \left[D_c-D\right]^{\beta},
\end{equation}
with $\beta =1/2$. The detailed derivation is provided in SM Sec.~S3.A. 
By anchoring the correlation length exponent $\nu=2$, we achieve an excellent data collapse for both $r$ and $U_4$. This collapse yields the order-parameter $r$ critical exponent to be $\beta=0.496\pm 0.011$. Notably, numerical estimation closely reproduces the analytical behavior, illustrating remarkable agreement with our numerical scaling analysis and robustly confirming the mean-field nature of the transition.

In the reminiscence of the macroscopic order-parameter $r$, it is quite important to examine the macroscopic behavior with diminishing the field-strength, \textit{i.e.,} $h\to 0$. In this limit, the continuous $U(1)$ rotational symmetry of the system is fully restored, and the onset of synchronization manifests as a spontaneous symmetry-breaking phase transition. The exact self-consistency equation simplifies to $r=\left.I_1\left(\frac{Kr}{D}\right)\right/I_0\left(\frac{Kr}{D}\right)$. The finite-size smearing of the order parameter directly informs the synchronization susceptibility, $\chi$, which links the macroscopic fluctuations to its linear response. In the thermodynamic limit, the isothermal susceptibility is defined by the response of the order parameter to a vanishing external field, which reads
\begin{equation}
    \chi = \left.\dfrac{\partial r}{\partial h}\right|_{h\to 0}.
\end{equation}
Furthermore, in the limit $h\to 0$ and $D>D_c$, we find
\begin{equation}
    \chi\approx [D-D_c]^{-1},
\end{equation}
indicating that near the critical noise-strength, it reveals that in the de-coherent phase ($D>D_c$), the susceptibility diverges as $\chi \sim |D-D_c|^{-1}$ (see Eq.~(S20) in SM). This establishes the mean-field critical exponent $\gamma=1$, governing a power-law divergence $\chi\propto |D-D_c|^{-\gamma}$. Crucially, at the zero-field ($h=0$) for a finite $N$, the divergent response is inherently truncated. Instead of relying on an external field, the finite-size susceptibility, $\chi_{_N}$, can be accurately extracted from the inherent thermal variance of the macroscopic order parameter
\begin{equation}
    \chi_{_N} = N(\langle r^2\rangle - \langle r\rangle^2).
\end{equation}
In the limit $N\to\infty$, the peak of $\chi_N$ becomes increasingly pronounced and asymptotically approaches the thermodynamic critical noise-strength $D_c=K/2$, as illustrated in Figs.~\ref{fig3} (a, b). According to the finite-size scaling theory, the maximum of this fluctuation-driven susceptibility scales as $\chi_{max} \propto N^{\gamma/\nu}$, where $\nu$ represents the finite-size scaling volume exponent. Utilizing the correlation length exponent $\nu = 2$ derived from our Binder cumulant collapse (yielding $\nu = 2$ for infinite-range interactions), we predict the peak height to scale precisely as $N^{1/2}$. 
Importantly, from the finite-size scaling analysis, we find $\gamma=1.064\pm 0.025$, indicating a good agreement between the theoretical and numerical estimations. Tracking $\chi_{_N}$, therefore, serves a dual purpose: it bypasses the need for an explicit external field while providing a robust, independent verification of the mean-field universality class underlying the spontaneous onset of synchronization.

An alternative strategy to investigate the synchronization transition can be obtained by a systematic study of the effective potential. The deterministic component of the coupled Langevin equations naturally descends from a many-body effective potential, which serves as an effective Hamiltonian for the ensemble. In the absence of thermal noise, this effective potential can be described as
\begin{equation}
    V(\{\theta_i\})=-h\sum_{i=1}^N\cos(\theta_i)-\dfrac{K}{2N}\sum_{i,j}\cos(\theta_i-\theta_j).
\end{equation}
Recalling the definition of macroscopic complex order parameter, \textit{i.e.,} $re^{\iota \psi}=\sum_{j=1}^Ne^{\iota \theta_j}/N$, the pairwise interactions gracefully decouple, and the effective potential is reduced to a purely macroscopic form
\begin{equation}
    V(\{\theta_i\})=-hNr-\dfrac{1}{2}KNr^2.
\end{equation}
In particular, we analyze the specific heat by exploring the energetic fluctuations associated with this effective potential in the strict zero-field limit ($h\to 0$). Without the external aligning field, the global phase $\psi$ is spontaneously selected, and the effective energy of the ensemble becomes directly proportional to the square of the order parameter.
With different sets of ensemble sizes, we recover the energy fluctuation at the criticality, which scales as $C\propto |D-D_c|^\alpha$. Moreover, numerically we get $\alpha=0.0276\pm 0.0119$ (see Figs.~\ref{fig3}(c, d)), which essentially recovers the thermodynamic signature that firmly locks the specific heat anomaly into the classical mean-field universality class.

\begin{figure}
    \centering
    \includegraphics[width=\linewidth]{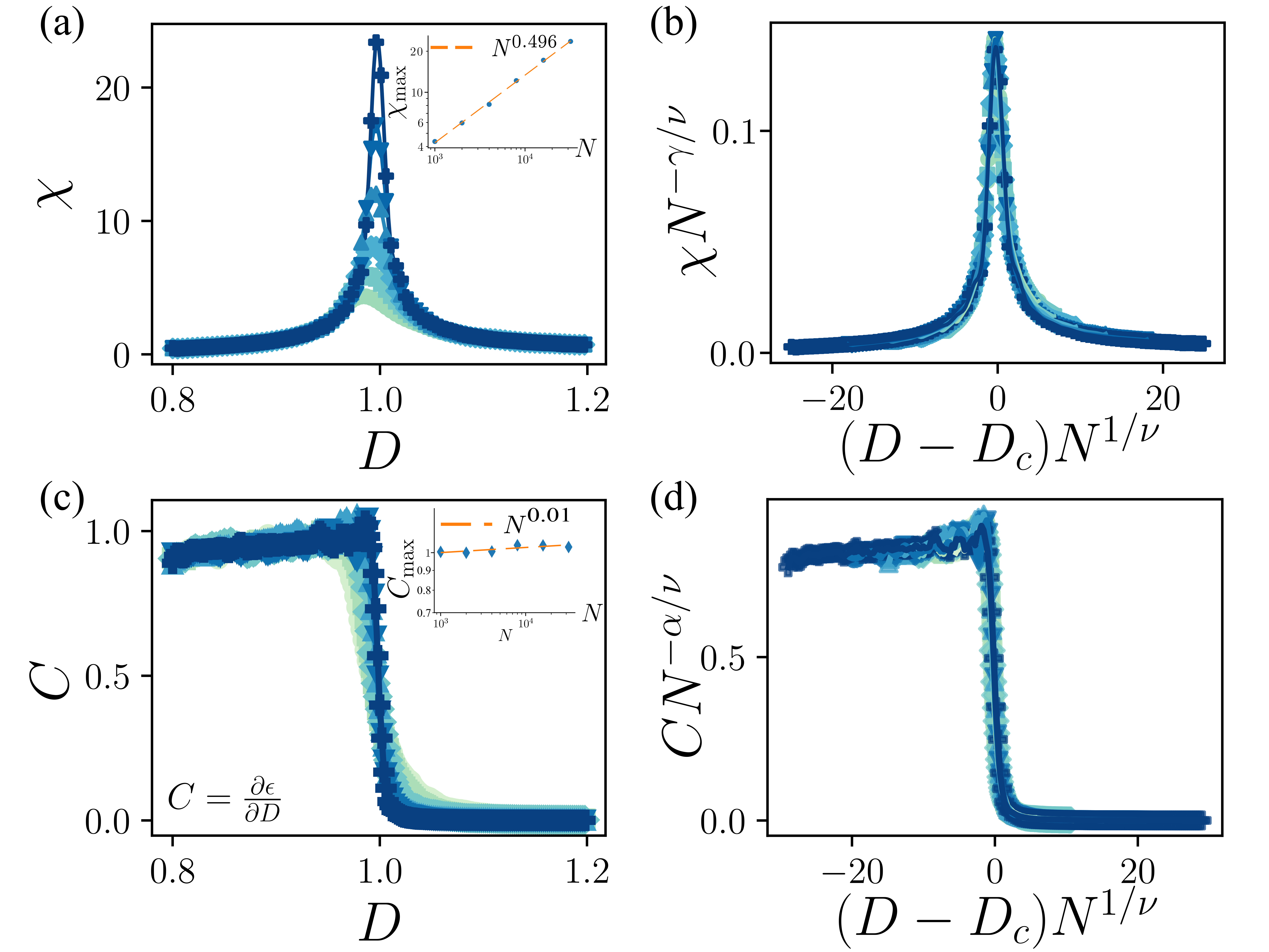}
    \caption{Thermodynamic fluctuation signatures and susceptibility scaling. Panel (a) illustrates susceptibility $\chi_N$ as a function of $D$, demonstrating that the peak becomes increasingly pronounced and approached to $D_c=K/2$ at $N\to \infty$. Panel (b) represents the data collapse of the fluctuation-driven susceptibility utilizing thr finite-size scaling volume exponent $\nu=2$ and the extracted exponent $\gamma=1.064\pm 0.025$. We describe the variation of the specific heat $C$ in pnel (c), in terms of the noise-strength $D$, which shows a finite discontinuity at $D_c$. In panel (d), we illustrate the data-collapse for the specific heat utilizing the scaling exponent $\alpha=0.0276\pm 0.0119$.}
    \label{fig3}
\end{figure}


\textbf{\textit{Conclusion}}-- In summary, we have systematically investigated the finite-size scaling behavior of globally coupled oscillators driven by thermal noise and a symmetry-breaking external field. By analytically deriving the exact stationary states via the Fokker-Planck formalism and employing rigorous finite-size scaling frameworks, we isolated the true thermodynamic critical point from the inherent finite-size smearing. Crucially, we extracted the critical exponents governing the order parameter $(\beta=0.496\pm 0.011)$, the synchronization susceptibility ($\gamma=1.064\pm 0.025$), and the specific heat anomaly ($\alpha=0.0276\pm 0.0119$).

The central achievement of this work is the definitive validation of the Rushbrooke inequality $\alpha+2\beta+\gamma \geq 2$, even within a non-equilibrium synchronization framework. This establishes a rigorous structural parallel, formal mapping the noise-induced synchronization transitions of finite dynamical networks onto the established universality classes of equilibrium critical phenomena. Looking forward, extending this exact scaling framework to more complex collective dynamics, such as swarmalators~\cite{o2017oscillators,pal2024directional,ghosh2024amplitude,sar2026interplay} with spatiotemporal coupling or non-Markovian systems governed by memory effects--presents a compelling avenue for uncovering new criticality. 
Finally, our investigation also calls for an entropic description of the synchronization process. This is because the model allows for 
a construction of a free energy function
\begin{equation}
{\cal {F}}[\rho]= - \int_{-\pi}^{\pi} (h + Kr) {\rm cos}\theta \rho(\theta,t) d\theta  - D \Sigma[\rho],
\end{equation}
having two distinct contributions. The first term here is  coming from the average internal energy, and the second is the entropic contribution with  $\Sigma(\rho) =  - \int_{-\pi}^{\pi} \rho(\theta, t) ~{\rm ln}\rho(\theta, t) d\theta$. It is easy to check that if $\rho(\theta, t)$ satisfies the Fokker-Planck equation (4),
\begin{equation}
\frac{d{\cal{F}}}{dt} = - \int_{-\pi}^{\pi} \rho(\theta, t) \Big[\frac{\partial}{\partial\theta}\Big( \frac{\delta {\cal{F}}}{\delta \rho}\Big)\Big]^2d\theta.
\label{eq:one}
\end{equation}
Note that since the integrand is positive, the free energy function decreases over time. Now, if the initial state is largely incoherent, the average internal energy contribution to $\cal{F}$ is
expected to decrease over time as the system moves towards synchronization. Further,  as the final state is more ordered, the Shannon entropy must decrease. However, to satisfy Eq. (\ref{eq:one}), the drop in Shannon entropy must be less than that in the first term, causing an overall reduction of ${\cal{F}}$. It would be worthwhile to check this for a finite-size system. The incorporation of entropy into the study of synchronization may open a new avenue of inquiry into synchronization phenomena from the perspective of non-equilibrium statistical mechanics.

\bibliography{main}

\newpage

\begin{titlepage}
    \centering
    {\Large \underline{Supplemental Material} \\[0.5cm]}
    {\Large \bf Thermodynamic criticality of coupled oscillators \\[0.5cm]}
    {\large Suvam Pal$^{1}$, Sudipta Mukherji$^{1}$, J\"urgen Kurths$^{2,3,4}$, Dibakar Ghosh$^{1}$ \\[0.3cm]}
    {\small 
        $^{1}$ Physics and Applied Mathematics Unit, Indian Statistical Institute, 203 B.T. Road, Kolkata, India
    }\\
    {\small 
        $^{2}$ Potsdam Institute for Climate Impact Research, Potsdam, Germany
    }\\
    {\small 
        $^{3}$ Department of Physics, Humboldt-Universit\"at zu Berlin, Berlin, Germany
    }\\
    {\small 
        $^{4}$ Research Institute of Intelligent Complex Systems, Fudan University, Shanghai, China
    }
\end{titlepage}

\onecolumngrid
\setcounter{page}{1}
\renewcommand{\thepage}{S\arabic{page}}

\setcounter{equation}{0}
\renewcommand{\theequation}{S\arabic{equation}}

\setcounter{figure}{0}
\renewcommand{\thefigure}{S\arabic{figure}}

\setcounter{section}{0}
\renewcommand{\thesection}{S\arabic{section}}

\setcounter{table}{0}
\renewcommand{\thetable}{S\arabic{table}}

\tableofcontents
\section{Derivation of the stationary state in Eq.~(\ref{fp})}
In this section, we provide the necessary steps to derive Eq.~\eqref{fp} in the main text. Recalling the Langevin equation from the main text (see Eq.~\eqref{thermo-dyn}). The Fokker-Planck equation reads
\begin{equation}\label{fp-supp}
\dfrac{\partial \rho(\theta,t)}{\partial t} = \dfrac{\partial}{\partial\theta} \left[ (h + K r)~\sin\theta~\rho(\theta,t)\right] + D\dfrac{\partial^2\rho(\theta,t)}{\partial\theta^2}.
\end{equation}
At $t\to \infty$, the configuration of the system can be captured with $\rho_{st}(\theta)$. In what follows, the left-hand side of Eq.~\eqref{fp-supp} is \textit{zero}.
\begin{equation}
\dfrac{\partial}{\partial\theta} \left[ (h + K r)\sin\theta \rho_{st}(\theta)\right] + D\dfrac{\partial^2\rho_{st}(\theta)}{\partial\theta^2} = 0.
\end{equation}
Integrating once, we get
\begin{equation}
  (h + K r)\sin\theta~\rho_{st}(\theta) + D\dfrac{\partial\rho_{st}(\theta)}{\partial\theta} = \mathcal{C}
\end{equation}
with $\mathcal{C}$  as an integration constant. It is now straightforward to argue that, to maintain the required periodicity of $\rho$ in $\theta$, $\mathcal{C}$ must be zero. For that, we notice that the equation has the form $y^\prime + p(x) y = q$. This equation is solved using an integrating factor. One multiplies both sides by $e^{\int p(x) dx}$ and then integrates. The result is
\begin{equation}
    y = e^{-\int p(x) dx} \int q e^{\int p(x)  dx} dx+ \mathcal{C}_1 e^{-\int p(x) dx},
\end{equation}
where $\mathcal{C}_1$ is a constant. In particular, we therefore get
\begin{equation}\label{one}
    \rho = \frac{C}{D} e^{\frac{h + K r}{D}{\rm cos}\theta} I + \mathcal{C}_1 e^{\frac{h + Kr}{D}{\rm cos}\theta},
\end{equation}
with the following identity
\begin{equation*}
    I = \int e^{-\frac{h+K r}{D} {\rm cos}\theta} d\theta.
\end{equation*}
Since the integrand is positive, the area under this curve always increases with $\theta$. Therefore
$I$ is not a periodic function. Hence, to maintain the periodicity of $\rho$, $\mathcal{C}$ in (\ref{one}) must be chosen to be zero.
In other words,
\begin{equation}
\rho_{st}(\theta) = \mathcal{C}_1 e^{\Big(\frac{h + Kr}{D}\Big)\cos\theta}.
\end{equation}
The unknown constant $\mathcal{C}_1$ is to be fixed by overall normalization of $\rho_{st}(\theta)$, namely using
$\int_0^{2\pi} \rho d\theta = 1$. Performing the integration, we reach at
\begin{equation}
    \mathcal{C}_1 = \frac{1}{2\pi I_0\Big[ \frac{h + K r}{D}\Big]}.
\end{equation}
Finally we get
\begin{equation}
    \rho_{st}(\theta) = \frac{e^{\Big(\frac{h + Kr}{D}\Big)\cos\theta}}{2\pi I_0\Big[ \frac{h + K r}{D}\Big]},
\end{equation}
which was reported in Eq.~(5) in the main text.

\section{Stationary state distribution in $D\to 0$}\label{s1}
Starting with the steady state distribution from the main text (see Eq.~(5)),
\begin{equation}\label{stationary-sup}
    \rho_{st}(\theta)=\dfrac{1}{2\pi}\dfrac{e^{\kappa \cos\theta}}{I_0(\kappa)},
\end{equation}
where $\kappa=(h+Kr)/D$. With $D\to 0$, which signifies $\kappa\to \infty$, we can write
\begin{equation}
    I_0(\kappa)\sim\dfrac{e^\kappa}{\sqrt{2\pi \kappa}},~~~~~~~~~~e^{\kappa \cos\theta}\sim e^{\kappa\left(1-\dfrac{1}{2}\theta^2\right)},\nonumber
\end{equation}
since, in the absence of noise, phases of the oscillators are distributed in the vicinity of $\theta=0$ (see Fig.~1(c)) with positive coupling strength $K>0$. Including the above relations, Eq.~\eqref{stationary-sup} reads
\begin{equation}\label{gaussian-supp}
    \rho_{st}(\theta)\simeq \sqrt{\dfrac{\kappa}{2\pi}} e^{-\dfrac{\kappa}{2}\theta^2},
\end{equation}
which is identically gaussian distribution with variance $\sigma^2=\kappa^{-1}$. Remarkably with $D\to 0$, $\kappa^{-1}\to 0$ which directly indicating $\sigma^2\to 0$. Furthermore, one can write $\rho_{st}\simeq\delta(\theta)$.

\section{Behavior of $r$, $\chi$ and $C$ near the criticality}

\subsection{Behavior of $r$ near the criticality}
The long-time behavior of the microscopic order parameter $r$ reads
\begin{equation}\label{supp-r1}
    r=\left.I_1\left(\dfrac{h+Kr}{D}\right)\right/I_0\left(\dfrac{h+Kr}{D}\right),
\end{equation}
where $I_0~\&~I_1$ denote the modified Bessel function of the first kind, as reported in Eq.~(6) in the main text. Since our primary analysis is focused on the behavior of the $r$ in the absence of an external field, by setting $h=0$, from Eq.~\eqref{supp-r1}, we get
\begin{equation}\label{supp-r2}
    r=\left.I_1\left(\dfrac{Kr}{D}\right)\right/I_0\left(\dfrac{Kr}{D}\right).
\end{equation}
Further, assuming $D_c$ to be the critical strength of noise, one can analytically derive the dependency of $r$ on $D_c$, by doing an expansion in the vicinity $r\to 0$. That reads
\begin{equation}\label{supp-low-r}
    r\simeq \frac{Kr}{2D}-\frac{K^3r^3}{16D^3},
\end{equation}
with the following solutions $r=0$ and $r=\pm\frac{4D}{K^{3/2}}\left[|D_c-D|\right]^{1/2}$, where $D_c=K/2$. Essentially, the non-zero positive root of Eq.~\eqref{supp-low-r} in $D\to D_c$, captures the behavior of $r$ at the criticality with exponent $\beta=1/2$. 

\subsection{Behavior of $\chi$ near the criticality}
We start by putting $x=(h+Kr)/D$ in Eq.~\eqref{supp-r1}, we get
\begin{equation}\label{rx}
    r = I_1(x)/I_0(x).
\end{equation}
By taking the derivative on both sides of the above identity with respect to $h$, we get
\begin{equation}\label{drdh-1}
    \dfrac{dr}{dh}=\dfrac{\partial}{\partial x}\left[\dfrac{I_1(x)}{I_0(x)}\right]\dfrac{dx}{dh},
\end{equation}
which can be further reduced with the following identity $\frac{dx}{dh} = 1+\frac{K}{D}\frac{dr}{dh}$, and we get
\begin{equation}\label{drdh-2}
    \dfrac{dr}{dh} = \dfrac{f'(x)}{D-Kf'(x)},
\end{equation}
where $f(x)=I_1(x)/I_0(x)$. Here $f'(x)=\left[I_1'(x)I_0(x)-I_0'(x)I_1(x)\right]/I_0(x)^2$. Further, using the recursive relations of the modified Bessel functions $I_0'(x)=I_1(x)$ and $I_1'(x)=I_0(x)-\frac{1}{x}I_1(x)$, from Eq.~\eqref{drdh-2} we get
\begin{equation}
    f'(x) = 1-\frac{1}{x}\frac{I_1(x)}{I_0(x)}-\frac{I_1(x)^2}{I_0(x)^2}.
\end{equation}
Essentially, using Eq.~\eqref{rx}, the above relation is reduced to be
\begin{equation}\label{fx-2}
    f'(x) = 1-\frac{r}{x}-r^2.
\end{equation}
Finally, using the above identity, from Eq.~\eqref{drdh-2}, we get
\begin{equation}\label{drdh-3}
    \dfrac{dr}{dh} = \dfrac{1-\frac{r}{x}-r^2}{D-K\left[1-\frac{r}{x}-r^2\right]}.
\end{equation}
In the regime $D>D_c$ with $h\to 0$, the existence of spontaneous order is not prominent, which essentially indicates $r\to 0$. Importantly, in this domain, the behavior of the susceptibility indicates the following identity
\begin{equation}
    \chi = \left.\dfrac{\partial r}{\partial h}\right|_{h\to 0} \simeq \dfrac{1}{2}\dfrac{1}{D-D_c},
\end{equation}
where $D_c = K/2$.

\subsection{Behavior of specific heat near criticality}
In the absence of an external field, the Lyapunov function (see Eq.~(12)), we get
\begin{equation}\label{supp-lyap}
    V=-\dfrac{1}{2}KNr^2.
\end{equation}
Note that $r$ is the intrinsic function of $K,~D$. Further differentiating Eq.~\eqref{supp-lyap} with respect to $D$, we get
\begin{equation}
    C=\dfrac{dV}{dD} = -KN~r~\dfrac{dr}{dD}.
\end{equation}
Further, recalling the behavior of $r$ in $D\to D_c$ limit, we find
\begin{equation*}
    \dfrac{dr}{dD}\simeq \dfrac{1}{2(D-D_c)^{1/2]}}.
\end{equation*}
Using the above identity, we find the behavior of $C$ at the critical $D$, which reads
\begin{equation}
    C\simeq (D-D_c)^0,
\end{equation}
which essentially identifies the behavior of the $C\propto (D-D_c)^\alpha$, with $\alpha=0$.





\end{document}